\documentstyle[preprint,aps]{revtex}

\def\text#1{{\hbox{\rm #1}}}
\begin{document}
\draft
\preprint{
\begin{tabular}{l}{\bf HU-SEFT R 1996-15}\\
{\bf AS-ITP-1996-11}\\
{\bf NCU/PHY/TP-960528}\\
{\bf hep-th/9605121}
\end{tabular}
}
\title{Differential Regularization of Topologically Massive 
Yang-Mills Theory and Chern-Simons Theory}
\author{W.F. Chen$^d$\renewcommand{\thefootnote}{\dagger}
\footnote{\small ICSC-World Laboratory, Switzerland}, 
H.C. Lee$^{e}$ and Z.Y. Zhu$^{a,b,c}$}

\address{${}^a$ CCAST (World Laboratory), P.O. Box 8730, 
Beijing, 100080, China\\
${}^b$ Institute of Theoretical Physics, Academia Sinica,\\
P. O. Box 2735, Beijing, 100080, China\\
${}^c$ Department of Physics, Chung-Hsing University, Taichung, Taiwan, 
China\\
${}^d$ Research Institute for High Energy Physics, University of Helsinki,\\
P.O. Box 9 (Siltavuorenpenger 20 C), FIN-000014, Helsinki, Finland\\
${}^e$ Department of Physics, National Central University,
Chungli, Taiwan 320, China
}

\maketitle

\begin{abstract}
\noindent
We apply differential renormalization method to the study of three-dimensional
topologically massive Yang-Mills and Chern-Simons theories. The method is
especially suitable for such theories as it avoids the need for dimensional
continuation of three-dimensional antisymmetric tensor and the Feynman
rules for three-dimensional theories in coordinate space are relatively
simple. The calculus involved is still lengthy but not as difficult as other
existing methods of calculation. We compute one-loop propagators and
vertices and derive the one-loop local effective action for 
topologically massive Yang-Mills theory. We then consider Chern-Simons
field theory as the large mass limit of topologically massive Yang-Mills
theory and show that this leads to the famous shift 
in the parameter $k$. Some useful formulas for the calculus of 
differential renormalization of
three-dimensional field theories are given in an Appendix.
\end{abstract}

\vspace{4mm}

\begin{flushleft}
{\it PACS: 03.70, 11.15. B, 11.10. G}\\
{\it Keywords: Differential regularization, Short-distance expansion, 
Local effective action, Finite renormalization}
\end{flushleft}
 


\vspace{3ex}

\section{INTRODUCTION}

The differential renormalization (DR) method was proposed by Freedman,
Johnson and Latorre${\cite{fjl}}$ to deal with the ultraviolet divergences
of quantum field theories few years ago. Its original idea came from the
observation that primitively divergent amplitudes are well defined in
coordinate space for non-coindent points, but too singular at short distance
to allow a Fourier transform into momentum space. They proposed to
renormalize such an amplitude by first writing its singular parts as
derivatives of some less singular functions that have well defined Fourier
transformations, then performing Fourier transformations of such
functions and discarding the surface terms. This idea is clearly illustrated
when one applies it to the one-loop 4-point bubble graph of 
massless $\phi^4$ theory in 4-dimensional space-time. 
As we know, the amplitude of this
graph involves the function $1/x^4$ that is singular at $x=0$, corresponding
to an ultraviolet divergence. To realize its differential renormalization we
follow ref.${\cite{fjl}}$ and use the identity, 
\begin{equation}
\displaystyle\frac 1{x^4}=
-\displaystyle\frac 14\Box \frac{\mbox{ln}(x^2M^2)}{x^2},
\quad {\hbox{\rm for}}~x{\neq }0.  \label{1}
\end{equation}
The function $\mbox{ln}(x^2M^2)/x^2$ has a well defined Fourier transform $4\pi
^2\ln (p^2/\overline{M}^2)/p^2$, where $\overline{M}=M/\gamma $ and $\gamma $
is Euler's constant. After discarding the surface term we are left 
with $-\pi ^2\ln (p^2/\overline{M}^2)$ as the regulated Fourier 
transform of $1/x^4$.

The DR method has been applied to many cases 
including massless ${\phi }^4$ theory up to three-loop 
order $\cite{fjl}$, one-loop massive ${\phi}^4$ theory $\cite{hala}$, 
supersymmetric Wess-Zumino model up to three loops $\cite{haa}$, Yang-Mills
theory in background field method up to one-loop $\cite{fjl}$, QED up to two
loops $\cite{hl}$ and low dimensional Abelian 
gauge theories to one-loop $\cite{ramon}$.
Its relation with the conventional dimensional regularization in some
theories $\cite{dn,sm}$ and compatibility with unitary have been
investigated $\cite{fjmv}$ and it has been shown to be simpler and more
powerful than other regularizations in many cases.

In this paper we shall use the DR method to study the perturbative three
dimensional topologically massive Yang-Mills theory (TMYM) and Chern-Simons
theory (CS) which, as will be shown, it is especially suited for.

The action of TMYM $\cite{djt}$, which is obtained by adding to the standard
non-Abelian gauge action, the Chern-Simons term, can be written in Euclidean
space as, 
\begin{equation}
\begin{array}{l}
S_m=-i\displaystyle \frac k{4\pi }\displaystyle\int_x {\epsilon }^{\mu \nu
\rho }\left(\displaystyle\frac 12A_\mu ^a{\partial }_\nu A_\rho ^a +\frac
1{3!} f^{abc}A_\mu ^aA_\nu ^bA_\rho ^c\right) +\displaystyle\frac{|k|}{
16m\pi ^2}\int_x F_{\mu \nu }^aF^{{\mu \nu }a}\, ,
\label{tmym}
\end{array}
\label{2}
\end{equation}
where the integration $\displaystyle \int_x \equiv \int d^3x$ 
is over the whole $R^3$. The
first term, i.e., the Chern-Simons term, exists only in three
dimensions. It is easy to see that under a gauge transformation $U$ the
action transforms as

\begin{eqnarray*}
S_m&\rightarrow&S_m-2{\pi}ikS_{WZ}\, , \\[2mm]
S_{WZ}&=&\displaystyle\frac{1}{24{\pi}^2}\int_x 
{\epsilon}^{\mu\nu\rho}\,\mbox{Tr}\left( U^{-1}{\partial}_{\mu}UU^{-1}
{\partial}_{\nu}UU^{-1} {\partial}_{\rho}U\right) \, .
\end{eqnarray*}
As it is well known, the Wess-Zumino term $S_{WZ}$ takes integer value, so
the theory is expected to be gauge invariant at the quantum level when $k$
takes integer value. At the same time, an interesting property of $S_m$ is
that the gauge excitations are massive, with mass $m$. This property exits
only in three dimensions and is not shared by other dimensions.

The perturbative property of TMYM was studied in $\cite{djt,prao}$, where it
was pointed out that the computations involved are not trivial and require
diligence. In our view, dimensionality plays an important role in defining 
three-dimensional TMYM because much of the topological properties of the
theory are derived from the properties of 
three dimensional antisymmetric tensor 
$\epsilon^{\mu \nu\rho }$. A calculation without using dimensional
continuation is therefore called for. We are furthermore motivated to study
the theory with DR by the fact that in three dimensions, propagators of TMYM
in coordinate space have analytic forms that are particularly suited for the
application of this method.

During the past several years a number of studies of perturbative
Chern-Simons theory have been carried out using a variety of
regularization schemes including: higher covariant derivative (HCD) combined
with generalized Pauli-Villars regularization $\cite{alr}$; HCD combined with
dimensional regularization $\cite{gmr1,csw}$; operator regularization $\cite
{cz}$; ${\eta }$ function regularization $\cite{brt1}$; geometric
regularization $\cite{asfa}$ and Feynman propagator
regularization $\cite{gmm}$. Especially recently an understanding on the
perturbative behaviour of CS from supersymmetric Yang-Mills-Chern-Simons 
theory has appeared $\cite{kll}$ and a more strict mathematical treatment
from the geometric viewpoint has been discussed in ref.$\cite{singer}$.
From these studies there emerges the so-called $k$-shift problem in
three-dimensional CS, which is concerned with whether quantum correction
change the value of the parameter $k$. 
It appears that whether the value of $k$ shifts or
not depends on the regularization scheme $\cite{asfa1,gmr2}$, 
{\em --} some of the
calculations in these studies showed the $k$-shift while others did not.
In ref.$\cite{asfa1}$, an analysis shows a family of shift can be
generated, which depends on the parity property of the regulator.

As we know, Chern-Simons action is just the first term in Eq.(\ref{tmym}).
Obviously we can consider TMYM as a partially (high covariant derivative)
regulated version of CS or, alternatively, CS as the large mass limit 
($m{\rightarrow }{\infty }$) of TMYM $\cite{gmr1}$. So a calculation of the
perturbative property of three-dimensional TMYM yields a study of the $k$
-shift problem of three-dimensional CS as a by-product. Our result confirms
the existence of $k$-shift and coincides with the case of scalar regulators 
of ref.$\cite{asfa1}$.

This paper is organized as follows. In section II we present the Feynman
rules of TMYM in coordinate space. Section III is devoted to explicit
calculations of one-loop amplitudes needed for the computation of 
one-loop local effective action, 
where we have obtained ghost self-energy, 
vacuum polarization tensor and gauge boson-ghost-ghost vertex. In
Section IV Slavnov-Taylor identity is explicitly derived and used in
combination of results from Section III to determine the one-loop local
effective action. In section V, as an example demonstrating the usefulness
of formulas given in Appendix, we give a result for the self-energy of
gauge field in three-dimensional QED. In section VI, we discuss and
summarize the results. Some formulas utilized in the calculation are given
in the Appendix. These formulas should also be useful for DR calculations of
other low-dimensional field theories.

\section{Feynman Rules in Coordinate Space}

By defining $g^2=4\pi /|k|$ and rescaling $A\rightarrow A/g$, we rewrite 
TMYM action (1) as 
\begin{equation}
\begin{array}{l}
S_m=-i\,\mbox{sgn}(k)\displaystyle\int_x {\epsilon }^{\mu \nu \rho } \left(%
\displaystyle \frac 12A_\mu ^a{\partial }_\nu A_\rho ^a+\frac
1{3!}gf^{abc}A_\mu ^aA_\nu ^bA_\rho ^c \right) +\displaystyle \frac{1}{4m}%
\int_x F_{\mu \nu }^aF^{{\mu \nu }a} \, ,
\end{array}
\label{3}
\end{equation}
whose corresponding BRST invariant action in the Landau gauge is 
\begin{equation}
S[A,c,\bar{c},B,m]=S_m+\int_x \left[{\partial }_\mu \bar{c}^aD^\mu c^a+B^a {%
\partial }_\mu A^{{\mu }a}\right].  \label{4}
\end{equation}
The BRST transformation of the fields are 
\begin{equation}
\begin{array}{l}
{\delta }A_\mu ^a=D_\mu c^a\, ,\qquad {\delta }\bar{c}^a=B^a \, , \\[2mm] 
\displaystyle{\delta }c^a=-\frac 12gf^{abc}c^bc^c\, ,\qquad {\delta }B^a=0\, .
\end{array}
\label{5}
\end{equation}
Here we choose the Landau gauge because of its good infrared behavior $\cite
{prao}$. For a pure Chern-Simons field theory, the Landau vector
supersymmetry $\cite{gmr2,brt2,dor}$, 
\begin{equation}
\begin{array}{rcl}
v_\mu A_\nu^a =i\,\mbox{sgn}(k)
{\epsilon }_{\nu \mu \rho }{\partial }^\rho c^a & , & 
v_\mu c^a=0 \, , \\[2mm] 
v_\mu {\bar{c}^a}=A_\mu^a & , & v_\mu B^a=-D_\mu c^a\, ,
\end{array}
\label{6}
\end{equation}
which only exists in the Landau gauge, plays a crucial role in the
cancellation of the infrared divergence. Although the inclusion of 
Yang-Mills term in TMYM breaks this symmetry, it does not ruin the
cancellation of the infrared singularity.

The generating functional can be formally written as 
\begin{equation}
Z[J,{\eta },\bar{\eta},M]=\displaystyle{\int}{\cal D}A{\cal D}B{\cal D}c 
{\cal D}\bar{c}\ \exp\left(-S - {\int}\left[J_\mu ^aA_\mu ^a+\bar{\eta}^ac^a+%
\bar{c}^a{\eta }^a + B^aM^a\right]\right) \, . \label{7}
\end{equation}
Differential regularization works in coordinate space, so we need the
Feynman rules in coordinate space. Defining 
\begin{eqnarray}
G_{\mu \nu }^{ab}(x-y) &=& <0|T[A_\mu ^a(x)A_\nu ^b(y)]|0>\, ,  \nonumber \\
{\Lambda}_\mu ^{ab}(x-y) &=& <0|T[A_\mu ^a(x)B^b(y)]0>\, ,  \nonumber \\
{\Lambda}^{ab}(x-y) &=& <0|T[B^a(x)B^b(y)]0>\, ,  \nonumber \\
S^{ab}(x-y) &=& <0|T[c^a(x){\bar{c}^b}(y)]|0>\, ,  \nonumber \\
<0|T[A_\mu ^a(x)A_\nu ^b(y)A_\rho ^c(y)]|0> &=& \displaystyle %
\int_{x^{\prime}} \int_{y^{\prime}} \int_{z^{\prime}} G_{\mu \mu ^{\prime
}}^{aa^{\prime }}(x-x^{\prime })G_{\nu \nu ^{\prime }}^{bb^{\prime
}}(y-y^{\prime })G_{\rho \rho ^{\prime }}^{cc^{\prime }}(z-z^{\prime }) 
 \nonumber \\
&&\times {\Gamma }_{(3){\mu}^{\prime}{\nu}^{\prime}
{\rho}^{\prime}}^{a^{\prime }b^{\prime }c^{\prime }}(x^{\prime
},y^{\prime },z^{\prime })\, ,  \nonumber \\
<0|T[A_\mu ^a(x)A_\nu ^b(y)A_\rho ^c(z)A_\sigma ^d(w)|0> &=& \displaystyle %
\int_{x^{\prime}} \int_{y^{\prime}} \int_{z^{\prime}} G_{\mu \mu
^{\prime}}^{aa^{\prime }}(x-x^{\prime }) G_{\nu \nu ^{\prime
}}^{bb^{\prime}}(y-y^{\prime }) \times  \nonumber \\
&& \hskip -8pt G_{\rho \rho ^{\prime }}^{cc^{\prime }}(z-z^{\prime
})G_{\sigma \sigma ^{\prime }}^{dd^{\prime }}(w-w^{\prime }) {\Gamma }%
_{(4)\mu ^{\prime}\nu ^{\prime }\rho ^{\prime } \sigma ^{\prime
}}^{a^{\prime }b^{\prime} c^{\prime }d^{\prime }}(x^{\prime },y^{\prime
},z^{\prime },w^{\prime })\, ,  \nonumber \\
<0|T[c^a(x){\bar{c}}^c(z)A_\mu ^b(y)]|0> &=& \displaystyle \int_{x^{\prime}}
\int_{y^{\prime}} \int_{z^{\prime}} S^{aa^{\prime }}(x-x^{\prime}) G_{\mu
\mu ^{\prime }}^{bb^{\prime }}(y-y^{\prime }) S^{c^{\prime}c}(z^{\prime }-z)
\nonumber \\
&& \times {\Lambda }_{\mu^{\prime}}^{a^{\prime }
b^{\prime }c^{\prime }}(x^{\prime
},y^{\prime },z^{\prime })\, ,  \label{8}
\end{eqnarray}
we obtain Feynman rules as follows (Fig. 1), 
\begin{eqnarray}
G_{\mu \nu }^{(0)ab}(x-y)&\equiv& {\delta }^{ab}D_{\mu \nu }(x-y)  \nonumber
\\
&=&-{\delta }^{ab} \left[i\,\mbox{sgn}(k){\epsilon }_{\mu \nu \rho }{\partial}
_x^\rho + \displaystyle\frac 1m \left({\delta}_{\mu \nu }{{\nabla}^2}^x-{
\partial }_\mu ^x {\partial }_\nu ^x\right) \right]\displaystyle\frac{
(1-e^{-m|x-y|})}{4\pi |x-y|}\, ,  \nonumber \\
{\Lambda}_\mu ^{(0)ab}(x-y) &\equiv& {\delta }^{ab}\Lambda _\mu (x-y)=-{
\delta }^{ab}{\partial }_\mu ^x\displaystyle\frac 1{4\pi |x-y|}\, ,  \nonumber
\\
{\Lambda}^{(0)ab}(x-y)&=&0\, ,  \nonumber \\
S^{(0)ab}(x-y) &\equiv& {\delta }^{ab}S(x-y) = {\delta }^{ab}\displaystyle
\frac 1{4\pi |x-y|}\, ,  \nonumber \\
{\Gamma }_{(3)\mu \nu \rho }^{(0)abc}(x,y,z)&=&gf^{abc}\left[i\,\mbox{sgn}(k){
\epsilon }_{\mu \nu \rho }- \displaystyle\frac 1m[({\partial }_\mu ^y-{
\partial }_\mu^z) {\delta }_{\nu \rho }+({\partial }_\nu ^z-{\partial }_\nu
^x) {\delta }_{\rho \mu } \right.  \nonumber \\
&+& \hskip -6pt \left.({\partial }_\rho ^x-{\partial }_\rho ^y){\delta }
_{\mu \nu } \right] \displaystyle \int_u {\delta }^{(3)}(x-u){\delta }
^{(3)}(y-u) {\delta }^{(3)}(z-u)\, ,  \nonumber \\
{\Gamma }_{(4)\mu \nu \rho \sigma }^{(0)abcd}(x,y,z,w)&=& \displaystyle\frac{
g^2}m \left[ f^{eab}f^{ecd}({\delta }_{\nu \rho }{\delta }_{\sigma \mu }- {
\delta }_{\nu \sigma }{\delta }_{\mu \rho }) +f^{eac}f^{edb}({\delta }_{\rho
\sigma } {\delta }_{\nu \mu }-{\delta }_{\nu \rho }{\delta }_{\mu \sigma
})\right.  \nonumber \\
+ f^{ead}f^{ebc}({\delta }_{\sigma \nu }{\delta }_{\rho \mu } \hskip -8pt
&-& \hskip-8pt \left.{\delta }_{\sigma \rho }{\delta }_{\mu \nu })\right] 
\displaystyle \int_u {\delta }^{(3)}(x-u) {\delta }^{(3)}(y-u){\delta }
^{(3)}(z-u){\delta }^{(3)}(w-u)\, ,  \nonumber \\
{\Lambda }_\mu ^{(0)abc}(x,y,z) &=& gf^{abc}{\partial }_\mu ^x\displaystyle
\int_u {\delta }^{(3)}(x-u){\delta }(y-u){\delta }(z-u)\, ,  \label{9}
\end{eqnarray}
where the superscript ``\,$(0)$\,'' denotes free propagators or bare vertices.
We can see that the propagators given above are much simpler in comparison
with their counterparts in 4-dimensional massive theories, which are Bessel
functions $\cite{bs}$.

\section{One-Loop Amplitudes}

Now we use DR to to carry out the one-loop renormalization of TMYM. Naive
power counting suggests that some of the one-loop diagrams should be
ultraviolet divergent. But as we will show, in some sense, TMYM is
essentially a finite theory $\cite{prao}$. Purely for the purpose of making
this finiteness manifest (DR does not require it), we introduce a short
distance cutoff by excluding a small ball $B_\epsilon$ of radius $\epsilon$
about the origin as in $\cite{fjl,fjmv}$. Denote the region $R^3-B_{\epsilon}
$ by ${R^3_\epsilon}$.

Let us analyze the one-loop ghost self-energy first. Its Feynman diagram is
shown in Fig. 2. The Fourier transformation of its amplitude is 
\begin{equation}
\displaystyle -g^2C_V{\delta}^{ab}~\int_{R^3_\epsilon}{\partial }_\mu
^x(e^{-ip.x})D_{\mu \nu }(x){\partial }_\nu ^xS(x)\, .  \label{10}
\end{equation}
Here we would like to emphasize that we need to be careful about the
positions of the partial differential operators in the Feynman rules of (9).
By using (9) and the formulas in Appendix we have 
\begin{equation}
S^{(1)ab}(p)=\displaystyle {\delta }^{ab}\frac{g^2C_V}{8\pi^2m}
\int_{R^3_\epsilon} e^{-ip.x} 
ip_\mu x_\mu \left[\frac{1-e^{-mr}}{r^6}-\frac{me^{-mr}}{r^5}
\right]\, .  \label{11}
\end{equation}
Writing singular functions at $r=0$ in the above integrand as derivatives of
less singular functions, we get 
\begin{eqnarray}
S^{(1)ab}(p) &=&\displaystyle -{\delta }^{ab}\frac{g^2C_V}{8\pi ^2m}
\int_{R^3_\epsilon} e^{-ip.x}
ip_\mu \partial _\mu \left[\nabla ^2\frac{1-e^{-mr}}
{8r^2} \right.  \nonumber \\
&&\left.+\displaystyle\frac{m^2e^{-mr}}{4r^2}-\frac{m^3e^{-mr}}{6r}- \frac{%
m^4{\hbox{\rm Ei}}(-mr)}8 \right]\, ,  \label{12}
\end{eqnarray}
where ${\hbox{\rm Ei}}(x)$ is the exponential integral function: 
\begin{equation}
{\hbox{\rm Ei}}(-mr)=\displaystyle\int_{-\infty }^{-mr}dt\;\frac{e^t}t \, .
\label{13}
\end{equation}
Particular attentions should be paid to differential operators in (12) when
we perform the Fourier transformation. For example, 
\begin{equation}
\begin{array}{l}
\displaystyle\int_{R^3_\epsilon} ~e^{-ip.x}\nabla ^2\frac{1-e^{-mr}}{8r^2}
=\int_{R^3_\epsilon} ~ \left[ \nabla ^2\left( e^{-ip.x}\frac{1-e^{-mr}}
{8r^2}\right)\right.\nonumber \\ 
\displaystyle \left. -2~\partial _\mu e^{-ip.x}\partial _\mu \frac{1-e^{-mr}%
}{8r^2} +p^2~e^{-ip.x}\frac{1-e^{-mr}}{8r^2} \right]\nonumber \\ 
=\displaystyle\frac{m\pi }2+p^2\displaystyle\int_{R^3_\epsilon} ~e^{-ip.x}%
\frac{1-e^{-mr}}{8r^2} \, ,
\end{array}
\label{14}
\end{equation}
where the first term on the right-hand-side is a surface term from the
cut-off ball $B_\epsilon $. Finally, using the formulas in Appendix, we
obtain the one-loop ghost self-energy 
\begin{equation}
\displaystyle S^{(1)ab}(p)=-{\delta }^{ab}\frac{g^2C_V}{16\pi^2}p^2 \left[%
\frac{\pi p}{2m}+\frac{m^2}{p^2}-1- \frac{m}{p}\left(\frac{p}{m} -\frac{m}{p}%
\right)^2\arctan \frac{p}{m}\right]\, ,  \label{15}
\end{equation}
where $p{\equiv}|p|$.

The one-loop gluon vacuum polarization part can be computed in a similar
way. The proper gluon self-energy is determined by gauge symmetry to have the
form: 
\begin{eqnarray}
{\Pi }_{\mu \nu }^{ab}(p) &=&\delta ^{ab}\left[ \Pi _{\text{o}\mu \nu }(p)+\Pi
_{\text{e}\mu \nu }(p)\right]   \nonumber \\
&=&\delta ^{ab}\left[\mbox{sgn}(p){\epsilon }_{\mu \nu \rho }p_\rho 
\Pi_{\mbox{o}}(p^2)-\frac 1m\left( {\delta }_{\mu \nu }p^2
-p_\mu p_\nu \right) \Pi_{\mbox{e}}(p^2)\right]\, ,  \label{16}
\end{eqnarray}
where the subscripts ``\,o\," and ``\,e\," denote parity-odd and 
parity-even respectively. The single ghost-loop contribution 
to the vacuum polarization tensor (Fig. 3b) is 
\begin{equation}
\displaystyle -{\delta }^{ab}\frac{g^2C_V}{16\pi ^2m}\,
\int_{R_\epsilon ^3}~e^{-ip.x}
\partial _\mu \frac 1r\partial _\nu \frac 1r \, .  \label{17}
\end{equation}
Combining it with the contribution from the singular gluon-loop, we have 
\begin{equation}
\begin{array}{rcl}
\Pi _{\text{o}\mu \nu }(p) & = & \displaystyle i\,\mbox{sgn}(k)
\frac{g^2C_V}{16\pi ^2} \int_{R_\epsilon ^3}e^{-ip.x}
{\epsilon }_{\mu \nu \rho }\partial ^\rho \left[
\frac 9{m^3r^6}(1-e^{-mr})^2\right.  \\[2mm]
&  & \displaystyle -\frac{18}{m^2r^5}(1-e^{-mr})e^{-mr}+\frac 1{mr^4}\left(
-1-\frac{13}2e^{-mr}+\frac{33}2e^{-2mr}\right) -\frac 1{2r^3}e^{-mr} \\[2mm]
&  & \displaystyle \left. +\frac 9{r^3}e^{-2mr}+\frac m{4r^2}e^{-mr}-\frac{
m^2}{4r}e^{-mr}-\frac{m^2}{4r}e^{-mr}-\frac{m^3}4{\hbox{\rm Ei}}(-mr)\right] 
\\ 
& = & \displaystyle i\,\mbox{sgn}(k)\frac{g^2C_V}{16\pi^2} 
\int_{R_\epsilon ^3}~e^{-ip.x}
{\epsilon }_{\mu \nu \rho }\partial ^\rho \displaystyle \left[ \frac
3{8m^3}(\nabla ^2)^2\frac{(1-e^{-mr})^2}{r^2} \right. \\[2mm]
& &\displaystyle +\nabla ^2\left( -\frac
12+\frac 54e^{-mr}-\frac 34\frac{e^{-2mr}}{mr^2}\right)    
+\displaystyle \frac m{r^2}\left(-\frac 14e^{-mr}-3e^{-2mr}\right)
\\[2mm]
& &\displaystyle \left. -\frac{m^2}{4r}e^{-mr}
-\frac{m^3}4{\hbox{\rm Ei}}(-mr)\right] \, .\label{18}
\end{array}
\end{equation}
Again after performing Fourier transformation we obtain 
\begin{eqnarray}
{\Pi }_{\text{o}}(p^2) &=&
\displaystyle\frac{g^2C_V}{16\pi ^2}\left[ \left( 3\frac{p^3}
{m^3}+5\frac pm-\frac mp-\frac{m^3}{p^3}\right) \arctan \frac pm+\left(
-\frac 32\frac{p^3}{m^3}\right. \right.   \nonumber \\
&&\left. \left. \displaystyle -3\frac pm+12\frac mp\right) \arctan \frac
p{2m}-\frac 34\pi \frac{p^3}{m^3}-\pi \frac pm+\frac{m^2}{p^2}+2\right]\, .
\label{19}
\end{eqnarray}
The calculation of one gluon-loop contribution to $\Pi_{\mbox{e}}(p^2)$ 
is similar but tedious. The result is 
\begin{eqnarray}
\Pi_{\mbox{e}}(p^2) &=&\int_{R_\epsilon ^3}e^{-ip.x}
\frac{g^2}{16{\pi }^2}C_V\left[ 
\displaystyle \frac 3{2m^4r^6}(1-e^{-mr})^2-\frac
3{m^3r^5}e^{-mr}(1-e^{-mr})\right.   \nonumber \\
&&+\displaystyle\frac 1{m^2r^4}(\frac 18-\frac{11}4e^{-mr}+\frac{33}%
8e^{-2mr})-\frac 1{mr^3}\left( \frac 74e^{-mr}+\frac{17}4\right) e^{-2mr} 
\nonumber \\
&&+\displaystyle\frac 1{r^2}\left( \frac 78e^{-mr}+\frac 14e^{-2mr}\right)
-\frac 1r\left( \frac m2e^{-2mr}-\frac{7m}8e^{-mr}\right)   \nonumber \\
&&-\displaystyle \left. m^2{\hbox{\rm Ei}}(-2mr)-\frac{7m^2}8{\hbox{\rm Ei}}%
(-mr)\right]   \nonumber \\
&=&\int_{R_\epsilon ^3}e^{-ip.x}\frac{g^2}{16{\pi }^2}C_V\left\{ {\nabla }%
^4\left[ \displaystyle\frac 1{16}\frac 1{m^4r^2}(1-e^{-mr})^2\right] \right. 
\nonumber \\
&&+\displaystyle{\nabla }^2\left[ \frac 1{m^2r^2}\left( \frac 1{16}-\frac
58e^{-mr}+\frac 9{16}e^{-2mr}\right) \right] +\frac 1{r^2}\left( \frac{13}%
8e^{-mr}-3e^{-2mr}\right)   \nonumber \\
&&-\displaystyle \left. \frac m{2r}e^{-2mr}-\frac{7m}{8r}e^{-mr}-m^2{%
\hbox{\rm Ei}}(-2mr)-\frac{7m^2}8{\hbox{\rm Ei}}(-mr)\right\}   \nonumber \\
&=&\displaystyle-\frac{g^2C_V}{32\pi }\left[ \left( -8\frac{m^3}{p^3}%
+24\frac mp+\frac 92\frac pm-\frac 12\frac{p^3}{m^3}\right) \arctan \frac
p{2m}+\right.   \nonumber \\
\left( -7{\frac{m^3}{p^3}}\right.  &-&\displaystyle \left. \left. 13\frac
mp-5\frac pm+\frac{p^3}{m^3}\right) \arctan \frac pm+11\frac{m^2}{p^2}+\frac
\pi 4\frac pm-\frac \pi 4\frac{p^3}{m^3}+5\right] \, .  \label{20}
\end{eqnarray}
The next step is to construct the local part of the one-loop effective
action and to demonstrate renormalization explicitly. From general
principles $\cite{lee}$ we know that this construction requires at least one
one-loop three-point Green function. Here we choose the one-loop vertex $Ac%
\bar{c}$, whose Feynman diagrams is shown in Fig. 4.

The amplitudes, which we know is divergentless from dimensional analysis,
can be written from Fig. 4 as 
\begin{equation}
V_\mu ^{abc}(p,q,r)=\displaystyle\frac
12g^3C_Vf^{abc}\int_x\int_ye^{-i(p.x+q.y)}\left[ V_\mu ^{(a)}(x,y)+V_\mu
^{(b)}(x,y)\right] \, ,  \label{21}
\end{equation}
where $p+q+r=0$. The contribution from Fig. 4a is 
\begin{equation}
\begin{array}{l}
V_\mu ^{(a)}(x,y)=iq_{\nu}{\partial }_\mu ^xS(x-y)
{\partial }_{\rho}^xS(x)D_{\nu \rho }(y) \\[2mm]
=-iq_{\nu }\displaystyle\frac 1{(4\pi )^3}{\partial }_\mu ^x\displaystyle
\frac 1{|x-y|}{\partial }_{\rho}^x\left[ \frac 1x\left( 1-e^{-mx}\right)
\right]  \\[2mm]
\times \left[-i\,\mbox{sgn}(k){\epsilon }_{\nu \rho\alpha }
{\partial }_\alpha ^y-
\displaystyle\frac 1m{\delta }_{\nu \rho}{{\nabla}^2}^y+\displaystyle\frac
1m{\partial }_{\nu}^y{\partial}_{\rho}^y\right]\displaystyle\frac 1y\left(
1-e^{-my}\right) \, .
\end{array}
\label{22}
\end{equation}
For our purpose, that is, to construct the local part of the effective
action, only the zero-momentum limit of this amplitude is needed, 
\begin{equation}
V_\mu ^{(a)abc}(p,q,r)=-g^3\frac{C_V}2f^{abc}\frac{17}{36}\frac 1{4\pi
}iq_\mu +\cdots \, .  \label{23}
\end{equation}
The amplitude from Fig. 4b can be reduced to 
\begin{equation}
\begin{array}{l}
V_\mu ^{(b)abc}(p,q,r)= \\[2mm]
\displaystyle\frac{C_V}2f^{abc}\int_x\int_y
e^{i(q.y+r.x)}\left[-i\,\mbox{sgn}(k){
\epsilon }_{\mu \nu\rho}q_{\lambda}{\partial }_{\sigma}^xS(x-y)D_{\nu
\lambda}(y)D_{\rho\sigma}(x)\right.  \\[2mm]
-\displaystyle\frac 1m p_{\rho}q_{\lambda}{\partial }_{\sigma}^xS(x-y)
D_{\mu\lambda}(y)D_{\rho\lambda}(x)+\displaystyle 
\frac 1m iq_{\lambda}{\partial }_{\sigma}^x S(x-y){\partial }_\mu ^y
D_{\rho\lambda}(y)D_{\rho\sigma}(x) \\[2mm]
-\displaystyle\frac 1m iq_{\lambda}{\partial }_{\sigma}^xS(x-y)
{\partial}_{\rho}^yD_{\mu\lambda}(y)D_{\rho\sigma}(x)
+\displaystyle\frac 1m iq_{\lambda}{\partial}_{\sigma
}^xS(x-y)D_{\nu\lambda}(y){\partial }_\mu ^xD_{\nu\sigma}(x) \\[2mm]
\left. -\displaystyle\frac 1m iq_{\lambda}{\partial }_{\sigma}^xS(x-y)D_{\nu
\lambda}(y){\partial }_{\nu}^xD_{\mu\sigma}(x)+\displaystyle\frac
1m p_{\nu}q_{\lambda}{\partial}_{\sigma}^xS(x-y)D_{\nu\lambda}(y)D_{\mu
\sigma}(x)\right] \, ,
\end{array}
\label{24}
\end{equation}
which, after a similar analysis and a lengthy calculation yields the
zero-momentum limit 
\begin{equation}
\begin{array}{l}
V_\mu ^{(b)abc}(p,q,r)=\displaystyle\frac{C_V}2f^{abc}\frac{17}{36}\frac
1{4\pi }iq_\mu +\cdots \, . 
\end{array}
\label{25}
\end{equation}
From Eqs.(21), (23) and (25), we conclude that one-loop $Ac\bar{c}$ vertex
takes the form 
\begin{equation}
V_\mu ^{abc}=0+\cdots \, .  \label{26}
\end{equation}
This means precisely that $\widetilde{Z}(0)=1$ to one-loop order, 
$\widetilde{Z}(0)$ denotes the $Ac\bar{c}$ vertex renormalization
constant defined at $p^2=0$. In fact
this is the correct result to any order in perturbation expansion for a
gauge theory.

\section{One-loop Local Effective Action}

Having computed the vacuum polarization tensor, the ghost self-energy and
the $Ac\bar{c}$ vertex, we are now in a position to derive the local
effective action. Our method is the same as that used in $\cite{gmr1}$ for
Chern-Simons theory.

We define the generating functional $Z[J,{\eta },\bar{\eta},M,K,L]$ with the
external fields $K_\mu ^a$ and $L^a$ respectively coupled to non-linear BRST
transformation products $D_\mu c^a$ and $-gf^{abc}c^bc^c/2$ as 
\begin{equation}
\begin{array}{l}
Z[J,{\eta },\bar{\eta},M,K,L]=\displaystyle{\int }{\cal D}X\exp \left[
-\left( S+\displaystyle{\int}_x\left( J_\mu ^aA^{\mu ^a}+\bar{\eta}^ac^a+%
\bar{c}^a{\eta }^a\right. \right. \right.  \\[2mm]
\left. \left. +B^aM^a+K_\mu ^aD^\mu c^a+L^a\left( -\displaystyle\frac 12
gf^{abc}c^bc^c\right) \right) \right] \, ,
\end{array}
\label{27}
\end{equation}
where $X=(A_\mu ,B,c,\bar{c})$. The Slavnov-Taylor identity arising from the
BRST transformation in (5) is 
\begin{equation}
\displaystyle\int_x\left[ J_\mu ^a\frac \delta {{\delta }K_\mu ^a}-\bar{\eta}%
^a\frac \delta {{\delta }L^a}+{\eta }^a\frac \delta {{\delta }M^a}\right]
Z=0\, .  \label{28}
\end{equation}
In addition, the invariance of $Z[J,{\eta },\bar{\eta},M,K,L]$ under the
translations $B^a(x)\to B^a(x)+{\lambda }^a(x),\;\bar{c}^a(x)\to \bar{c}%
^a(x)+{\omega }^a(x)$ leads respectively to the $B$-field and anti-ghost
field equations: 
\begin{equation}
\left[ {\partial }_\mu \displaystyle\frac \delta {{\delta }J_\mu^a}
-M^a\right] Z=0\, ,  \label{29}
\end{equation}
\begin{equation}
\left[ {\partial }_\mu \displaystyle\frac \delta {{\delta }K_\mu ^a}-{\eta}
^a\right] Z=0\, .  \label{30}
\end{equation}
Defining the generating functional for the connected Green function $W$ and
that for the one-particle-irreducible Green function $\Gamma $ ( i.e.,
the quantum effective action) as, 
\begin{equation}
\begin{array}{l}
W[J,{\eta },\bar{\eta},M,K,L]=-\mbox{ln}Z[J,{\eta },\bar{\eta},M,K,L] \\[2mm]
{\Gamma }[A_\mu ^a,B^a,c^a,\bar{c}^a,K_\mu ^a,L^a]=W[A_\mu ^a,B^a,c^a,\bar{c
}^a,K_\mu ^a,L^a]-\left( A^{\mu a}J_\mu ^a+B^aM^a+\bar{\eta}^ac^a+\bar{c}^a{
\eta }^a\right) \, ,
\end{array}
\label{31}
\end{equation}
we obtain the actions of the Slavnov-Taylor identity, the $B$-field and the
anti-ghost field equations on $\Gamma $: 
\begin{equation}
{\partial }_\mu A^{{\mu }a}+\displaystyle\frac{{\delta }{\Gamma }}{{\delta }%
B^a}=0\, ,  \label{32}
\end{equation}
\begin{equation}
{\partial }_\mu \displaystyle\frac{\delta {\Gamma }}{{\delta }K_\mu ^a}-%
\displaystyle\frac{\delta {\Gamma }}{{\delta }\bar{c}^a}=0\, ,  \label{33}
\end{equation}
\begin{equation}
\displaystyle\int_x\left[ \frac{{\delta }{\Gamma }}{{\delta }A^{\mu a}}\frac{%
{\delta }{\Gamma }}{{\delta }K_\mu ^a}-\frac{{\delta }{\Gamma }}{{\delta }c^a%
}\frac{{\delta }{\Gamma }}{{\delta }L^a}\right] =0\, .  \label{34}
\end{equation}
By a re-definition of $\Gamma $: 
\begin{equation}
\bar{\Gamma}=\Gamma +\displaystyle\int_xB^a{\partial }^\mu A_\mu ^a\, ,
\label{35}
\end{equation}
these equations become: 
\begin{equation}
\displaystyle\frac{{\delta }\bar{\Gamma}}{{\delta }B^a}=0,\qquad {\partial }%
_\mu \displaystyle\frac{\delta \bar{\Gamma}}{{\delta }K_\mu ^a}-%
\displaystyle \frac{\delta \bar{\Gamma}}{{\delta }\bar{c}^a}=0\, ,  \label{36}
\end{equation}
\begin{equation}
\displaystyle\int_x\left[ \frac{{\delta }\bar{\Gamma}}{{\delta }A^{\mu a}}%
\frac{{\delta }\bar{\Gamma}}{{\delta }K_\mu ^a}-\frac{{\delta }\bar{\Gamma}}{%
{\delta }c^a}\frac{{\delta }\bar{\Gamma}}{{\delta }L^a}\right] =0\, .
\label{37}
\end{equation}
The first relation in Eq.(36) means that the re-defined action $\bar{\Gamma}$
is independent of $B^a$ and the second relation implies that $K_\mu ^a$ 
and $\bar{c}^a$ always appear in $\bar{\Gamma}$ through the combination 
\begin{equation}
G_\mu ^a(x)=K_\mu ^a-{\partial }_\mu \bar{c}^a\, .  \label{38}
\end{equation}
Now we introduce the loop-wise expansion for $\bar{\Gamma}$: 
\begin{equation}
\bar{\Gamma}=\sum_{n=0}^\infty {\hbar }^n\bar{\Gamma}^{(n)}\, ,  \label{39}
\end{equation}
where $\bar{\Gamma}^{(0)}$ is the classical effective action without the
gauge-fixing term $\displaystyle\int_xB^a{\partial }^\mu A_\mu ^a$: 
\begin{equation}
\begin{array}{l}
\bar{\Gamma}^{(0)}=-i\,\mbox{sgn}(k)\displaystyle\int_x{\epsilon }^{\mu\nu\rho
}\left( \displaystyle\frac 12A_\mu ^a{\partial }_\nu A_\rho ^a+\frac
1{3!}gf^{abc}A_\mu ^aA_\nu ^bA_\rho ^c\right)  \\[2mm]
+\displaystyle\frac 1{4m}\int_xF_{\mu \nu }^aF^{{\mu \nu }a}+\displaystyle
\int_x\left[ G_\mu ^aD^{\mu}c^a+L^a\left( -\displaystyle\frac
12gf^{abc}c^bc^c\right) \right]\, .
\end{array}
\label{40}
\end{equation}
Substituting this expansion into Eq.(37) and comparing the coefficients of
the ${\hbar }^0$ and ${\hbar }^1$ terms lead to 
\begin{equation}
\displaystyle\int_x\left[ \frac{{\delta }{\Gamma }^{(0)}}{{\delta }A^{\mu a}}%
\frac{{\delta }\Gamma ^{(0)}}{{\delta }G_\mu ^a}-\frac{{\delta }{\Gamma }%
^{(0)}}{{\delta }c^a}\frac{{\delta }{\Gamma }^{(0)}}{{\delta }L^a}\right] =0
\label{41}
\end{equation}
and 
\begin{equation}
{\Delta }{\Gamma }^{(1)}=0\, ,  \label{42}
\end{equation}
where we have used the relation 
\begin{equation}
\displaystyle\frac{{\delta }}{{\delta }G_\mu ^a}=\displaystyle\frac{{\delta}
}{{\delta }K_\mu ^a}\, .  \label{43}
\end{equation}
${\Delta }$, the linear Slavnov-Taylor operator 
\begin{equation}
{\Delta }=\displaystyle\int_x\left[ \frac{{\delta }\bar{\Gamma}^{(0)}}{{
\delta }A^{\mu a}}\frac{{\delta }}{{\delta }G_\mu ^a}+\frac{{\delta }}{{
\delta }A^{\mu ^a}}\frac{{\delta }\bar{\Gamma}^{(0)}}{{\delta }G_\mu ^a}-
\frac{{\delta }\bar{\Gamma}^{(0)}}{{\delta }c^a}\frac{{\delta }}{{\delta }L^a
}-\frac{{\delta }}{{\delta }c^a}\frac{{\delta }\bar{\Gamma}^{(0)}}{{\delta }
L^a}\right] \, , \label{44}
\end{equation}
is the quantum analogue of the classical BRST operator and is nilpotent : 
\begin{equation}
{\Delta }^2=0\, .  \label{45}
\end{equation}
Now we follow the method of $\cite{gmr1,gmr2}$ to find the solution to
Eq.(42). From the requirement of zero ghost-number and mass dimension 3 we
determine the general form of one-loop effective action to be that${}^{*}$
\footnote{Rigorously speaking, the one-loop local effective action given
here is not perfect, it should contain other $(1/m)^n$ ($n{\geq}1$) dependent
higher covariant derivative terms such as $\displaystyle \frac{1}{m^{2n+1}}
{\int}_x F_{\mu\nu}(D^2)^nF^{\mu\nu}$, 
$\displaystyle \frac{1}{m}{\Delta} {\int}_x{\epsilon}^{\mu\nu\rho}G_{\mu}^a$
and $\displaystyle \frac{1}{m}{\Delta}{\int}_x (L^ac^a) (\bar{c}^bc^b)$ etc.
They also have correct mass dimension and ghost number. However
since in this section our aim is at the large $m$-limit, we have put these
$1/m$ terms out of consideration}
\begin{eqnarray}
{\Gamma }^{(1)} &=&{\alpha }_1\left[-i\,\mbox{sgn}(k)
\displaystyle\int_x~{\epsilon }
^{\mu \nu \rho }\left( \displaystyle\frac 12A_\mu ^a{\partial }_\nu A_\rho
^a+\frac 1{3!}gf^{abc}A_\mu ^aA_\nu ^bA_\rho ^c\right) \right]   \nonumber \\
&&+{\alpha }_2\displaystyle\frac 1{4m}\int_xF_{\mu \nu }^aF^{{\mu \nu }a}+{
\Delta }\displaystyle\int_x~[{\beta }_1G^{\mu a}A_\mu ^a+{\beta }_2L^ac^a] 
\nonumber \\
&=&-i\,\mbox{sgn}(k)({\alpha }_1+2{\beta }_1)
\displaystyle\int_x\frac 12{\epsilon }
^{\mu \nu \rho }A_\mu ^a{\partial }_\nu A_\rho ^a  \nonumber \\
&&-i\,\mbox{sgn}(k)({\alpha }_1+3{\beta }_1)\int_x\frac 1{3!}
{\epsilon }^{\mu \nu \rho }f^{abc}A_\mu ^aA_\nu ^bA_\rho ^c  \nonumber \\
&&+({\alpha }_2+2{\beta }_1)\displaystyle\frac 1{4m}\int_x({\partial }^\mu
A^{\nu a}-{\partial }^\nu A^{\mu a})({\partial }_\mu A_\nu ^a-{\partial }
_\nu A_\mu ^a)  \nonumber \\
&&+({\alpha }_2+3{\beta }_1)\displaystyle\frac 1{2m}\int_xgf^{abc}({\partial 
}_\mu A_\nu ^a-{\partial }_\nu A_\mu ^a)A^{\mu b}A^{\nu c}  \nonumber \\
&&+({\alpha }_2+4{\beta }_1)\displaystyle\frac
1{4m}\int_xg^2f^{eab}f^{ecd}A_\mu ^aA_\nu ^bA^{\mu c}A^{\nu d}+{\beta }_1
\displaystyle\int_xG_\mu ^a{\partial }^\mu c^a  \nonumber \\
&&+{\beta }_2\displaystyle\int_xG_\mu ^aD^\mu c^a-{\beta }_2
\displaystyle \int_x\frac 12gf^{abc}c^bc^c\, ,  \label{46}
\end{eqnarray}
where ${\alpha }_i$ and ${\beta }_i$ are constant coefficients. In
comparison with CS $\cite{gmr1,gmr2}$, we find that the  formal
large-$m$ limit of above effective action 
has the same form as that in refs.$\cite{gmr1,gmr2}$, 
i.e., the difference lies only in the mass dependent terms. By
using the results given in the last section and choosing the
renormalization point at $|p|=0$,  we can determine the values of the
parameters as follows: 
\begin{equation}
{\alpha }_1=\displaystyle\frac{g^2C_V}{4\pi },\quad {\alpha }_2=-
\displaystyle \frac{g^2C_V}{32\pi },\quad {\beta }_1=-\displaystyle\frac{
g^2C_V}{16\pi },\quad {\beta }_2=0\, .  \label{47}
\end{equation}
Thus, up to one-loop the explicit local effective action is: 
\begin{eqnarray}
{\Gamma }_{\mbox{local}}&=&{\Gamma }^{(0)}
+{\Gamma }_{\mbox{local}}^{(1)}  \nonumber \\
&=&\displaystyle\left( 1+\frac 1{4\pi }g^2C_V\right)
\left[-i\,\mbox{sgn}(k)\right] 
\displaystyle\int_x{\epsilon }^{\mu \nu \rho }\left( \frac 12A_\mu ^a{
\partial }_\nu A_\rho ^a+\frac 1{3!}A_\mu ^aA_\nu ^bA_\rho ^c\right)  
\nonumber \\
&&+\displaystyle\frac 1{4m}\left( 1-\frac 1{32\pi }g^2C_V\right) 
\displaystyle\int_xF^{\mu \nu a}F_{\mu \nu }^a  \nonumber \\
&&-{\Delta }\displaystyle \left( \frac 1{16\pi }g^2C_V\int_xG_\mu ^aA_\mu
^a+\int_xL^ac^a\right) +\displaystyle\int_xB^a{\partial }^\mu A_\mu ^a 
\nonumber \\
&=&\left( 1+\displaystyle\frac 1{4\pi }g^2C_V\right)
 \left[-i\,\mbox{sgn}(k)\right] 
\displaystyle\int_x{\epsilon }^{\mu \nu \rho }\left( \frac 12A_\mu ^a{
\partial }_\nu A_\rho ^a+\frac 1{3!}A_\mu ^aA_\nu ^bA_\rho ^c\right)  
\nonumber \\
&&+\displaystyle \frac 1{4m}\displaystyle\left( 1+\frac 9{32\pi }{g^2C_V}%
\right) ^{-1}\int_xF^{\mu \nu a}F_{\mu \nu }^a  \nonumber \\
&&-\displaystyle {\Delta }\left( \frac 1{16\pi }g^2C_V\int_xG^{\mu a}A_\mu
^a+\int_xL^ac^a\right) +\displaystyle\int_xB^a{\partial }^\mu A_\mu ^a\, .
\label{48}
\end{eqnarray}
Finally the one-loop effective action of CS can be easily
obtained by taking the large-mass limit $m{\rightarrow }\infty $. Obviously
the wave-function renormalization constants are 
\begin{equation}
Z_A=Z_B^{-1}=Z_G^{-1}=1-\displaystyle\frac 1{16\pi }{g^2C_V},\qquad
Z_L=Z_C^{-1}=1\, .  \label{49}
\end{equation}
This result can be cast into $k-$shift form, i.e., 
\begin{equation}
k\to k+\mbox{sgn}(k)C_V\, .  \label{50}
\end{equation}

\section{Gauge Field Self-Energy in Three-Dimensional QED}

In the Appendix are given some formulas which are useful for computation in
the study of any three-dimensional theory in coordinate space. Here, by the
way, we would like to point out that, as an example, using these formulas we
can get the one-loop self-energy part for the gauge field in
three-dimensional massive QED an analytic expression whose integral form was
given in ref.$\cite{ramon}$, 
\begin{eqnarray}
\Pi _{ij}(p) &=&-\frac{e^2}{8\pi }\left( \delta _{ij}p^2-p_ip_j\right)
\left[ \frac{2m}{p^2}+\left( \frac 1p-\frac{4m^2}{p^3}\right) \arctan \frac
p{2m}\right]   \nonumber \\
&&-me^2\epsilon _{ijk}\frac 1{2\pi p}\arctan \frac p{2m}\, ,
\end{eqnarray}
where the notation is the same as that in ref.$\cite{ramon}$.

\section{Summary}

We carried out the one-loop calculation of the topologically massive
Yang-Mills theory and Chern-Simons theory in coordinate space using the
method of differential renormalzation. Our calculation shows that the method
is very powerful and is especially suited for quantum field theories in
three dimensions. The results we obtained 
on TMYM coincide with those of ref.$\cite{prao}$, which used the 
method of dimensional regularization. However,
as was pointed out in ref.$\cite{prao}$, the calculus of dimensional
regularization for a theory in three dimensions is subtle and perhaps even
problematic, not least because of the need for a dimensional continuation of
the antisymmetric tensor ${\epsilon }_{\mu \nu \rho }$; it is not known to
what extent the calculated renormalization of a field theory such as
TMYM, whose property is closely tied to the dimension of space-time, could be
an artifact of this continuation. In differential renormalzation there is
not such an ambiguity because one does not change the dimension of
space-time so there is no need for a continuation of the antisymmetric
tensor. It is therefore reassuring that the two sets of results agree.
For Chern-Simons field theory our result shows the shift $k$ to
$k+\text{sgn}(k)C_V$, which coincides with the case of scalar regulator of 
ref.$\cite{asfa1}$.

\vspace{1cm}

\noindent{\bf Acknowledgment:} \\
{\noindent
WFC is grateful to the World
Laboratory, Switzerland for financial support. He would also like 
to thank Professor Masud
Chaichian for his encouragements and useful discussions and the
Research Institute for High Energy Physics, University of
Helsinki for warm hospitality. HCL is partly supported by the
NSC(ROC) research grant 85-2112-M-008-017. 
ZYZ is partially supported by NFC of China
and LWTZ-1298 of Chinese Academy of Sciences. 
We are greatly indebted to Professor Manuel Asorey for explaining us their
results and pointing out some inaprropriate remarks in the orginal manuscript.
We would also like to thank Dr. Alberto Accardi
for reminding of us the paper by E.R. Bezerra De Mello. }

\eject

\appendix
\setcounter{equation}{0}

\section{}

\subsection{Differential formulas}

Defining

\[
f^{(n)}\equiv \left( \frac 1r\frac d{dr}\right)^ nf(r)\, ,\qquad f(r)=\frac{%
1-e^{-mr}}r\, , 
\]
\stepcounter{equation}\vskip -25pt {\hfill (\Alph{section}\arabic{equation})%
\break} \label{A1} \vskip 13pt\noindent
and denoting ${\partial }_1={\partial }_{i_1}$, $x_1=x_{i_1}$ etc, we have 
\begin{eqnarray*}
\partial _1\partial _2\cdots\partial _{2n}f &=& (\delta _{12}\delta
_{34}\cdots \delta_{2n-1,2n}+{\hbox{\rm  permutations}})f^{(n)}  \nonumber \\
&&+ (x_1x_2\delta _{34}\delta _{56}+{\hbox{\rm permutations}})f^{(n+1)} 
\nonumber \\
&& +(x_1x_2x_3x_4\delta _{56}\delta _{78}\cdots \delta _{2n-1,2n} +{%
\hbox{\rm permutations}})f^{(n+2)}  \nonumber \\
&&+ \cdots +(x_1x_2\cdots x_{2n})f^{(2n)}\, ,
\end{eqnarray*}
\stepcounter{equation}\vskip -25pt {\hfill (\Alph{section}\arabic{equation})%
\break} \label{A2}

\begin{eqnarray*}
\partial _1\partial _2...\partial _{2n+1} f &=& (x_1\delta
_{23}\delta_{45}\cdots \delta _{2n,2n+1} +{\hbox{\rm permutations}})f^{(n+1)}
\nonumber \\
&& +(x_1x_2x_3\delta _{45}\cdots \delta _{2n,2n+1} +{\hbox{\rm permutations}}%
)f^{(n+2)}  \nonumber \\
&& +\cdots +(x_1x_2\cdots x_{2n+1})f^{(2n+1)}\, .
\end{eqnarray*}
\stepcounter{equation}\vskip -25pt {\hfill (\Alph{section}\arabic{equation})%
\break}\label{A3} \vskip 13pt\noindent 
Some examples are: 
\begin{eqnarray*}
\partial _if &=& x_if^{(1)}\, ,  \nonumber \\
\partial _i\partial _jf &=& \delta _{ij}f^{(1)}+x_ix_jf^{(2)}\, ,  \nonumber \\
\partial _i\partial _j\partial _kf &=& (x_i\delta _{jk}+x_j\delta
_{ki}+x_k\delta _{ij})f^{(2)}+x_ix_jx_kf^{(3)}\, .
\end{eqnarray*}
\stepcounter{equation}\vskip -25pt {\hfill (\Alph{section}\arabic{equation})%
\break}\label{A4}

\subsection{Fourier transforms of some functions}

\begin{eqnarray*}
\displaystyle\int_x \frac{1}{x^2}e^{ip.x} &=& \displaystyle\frac{
2{\pi}^2}{|p|}\, .  \nonumber \\
\displaystyle\int_x\frac{e^{-nmx}}{x}e^{ip.x} &=& \displaystyle 
\frac{4\pi}{n^2m^2+p^2}\, .  \nonumber \\
\displaystyle\int_x \frac{e^{-nmx}}{x^2}e^{ip.x} &=& 
\displaystyle\frac{4\pi}{p}\arctan\frac{p}{nm}\, .  \nonumber \\
\displaystyle\int_x\text{Ei}(-nmx)
e^{ip.x} &=& -\displaystyle\frac{8{\pi}}{p^3} 
\left[\frac{\pi}{4}-\frac{1}{2} \arctan\displaystyle\frac{mn}{p}- 
\frac{1}{4}\frac{mnp}{p^2+m^2n^2}\right]\, .
\end{eqnarray*}
\stepcounter{equation}\vskip -25pt {\hfill (\Alph{section}\arabic{equation})%
\break}\label{A5}

\subsection{Integrals over $R^3_\epsilon$}

Recall that $R^3_\epsilon$ is $R^3$ excluding a small ball $B_\epsilon$ of
radius $\epsilon$ about the origin. 
\[
\displaystyle\int_{R^3_\epsilon}e^{-ip.x}{\partial }_\mu f(x)  =
-if(\epsilon )4{\pi }\displaystyle\frac{p_\mu }p\frac d{dp} \left[ \frac{
\sin(p\epsilon )}p\right] +ip_\mu \int_x e^{-ip.x}f(x)\, . 
\]
\stepcounter{equation}\vskip -25pt {\hfill (\Alph{section}\arabic{equation})
\break}\label{A6}

\begin{eqnarray*}
\displaystyle\int_{R^3_\epsilon} e^{-ip.x}{\partial }_\mu {\partial }
_\nu f(x) &=& 4{\pi }\displaystyle\frac 1\epsilon \frac
d{dx}f(x)|_{x=\epsilon } \displaystyle\frac \partial {{\partial }{p_\mu }}
\frac \partial {{\partial }{p_\nu }} \left[ \frac{\sin(p\epsilon )}p\right] 
\nonumber \\
&& +\displaystyle 4{\pi }f({\epsilon }) \displaystyle\frac \partial {{
\partial }{p_\mu }}\displaystyle \frac \partial {{\partial }{p_\nu }}\left[ 
\frac{\sin(p\epsilon )}p\right] -p_\mu p_\nu \displaystyle\int_x 
e^{-ip.x}f(x)\, .
\end{eqnarray*}
\stepcounter{equation}\vskip -25pt {\hfill (\Alph{section}\arabic{equation})
\break}\label{A7}

\begin{eqnarray*}
\displaystyle\int_{R^3_\epsilon} e^{-ip.x}{\partial }^2 {\partial }
_\mu f(x) &=& -i(4\pi ){\partial }^2f(x)|_{x={\epsilon }} \displaystyle\frac
\partial {{\partial }p_\mu } \left[ \frac{\sin(p\epsilon )}p\right]  \nonumber
\\
&& +\displaystyle i 4{\pi }p_\mu p_\alpha {\epsilon }f({\epsilon }) 
\displaystyle \frac \partial {{\partial }p_\alpha }\left[ \frac{%
\sin(p\epsilon )}p\right]  \nonumber \\
&&- i 4{\pi }{\epsilon } p_\mu \frac d{dx}f(x)|_{x=\epsilon } \frac{
\sin(p\epsilon )}p -ip_\mu p^2\int_x e^{-ip.x}f(x).
\end{eqnarray*}
\stepcounter{equation}\vskip -25pt {\hfill (\Alph{section}\arabic{equation})%
\break}\label{A8}

\begin{eqnarray*}
\displaystyle\int_{R^3_\epsilon} e^{-ip.x}{\partial }_\mu {\partial }_\nu {
\partial }_\rho f(x) &=& i4{\pi }{\delta }_{\nu \rho }\displaystyle\frac
\partial {{\partial }p_\mu }\frac{\sin(p\epsilon )}{p\epsilon } \left[ \frac
1x\frac d{dx}f(x)\right]_{x=\epsilon }  \nonumber \\
&& +\displaystyle i4{\pi } \frac{{\partial }^3}{{\partial }p_\mu {\partial }
p_\nu {\partial }p_\rho } \frac{\sin(p\epsilon )}{p\epsilon }\frac 1x \frac
d{dx}\left[ \frac 1x\frac d{dx}f(x)\right]_{x=\epsilon }  \nonumber \\
&& -ip_\mu \displaystyle{\int }_x e^{-ip.x}{\partial }_\nu 
{\partial }_\rho f(x).
\end{eqnarray*}
\stepcounter{equation}\vskip -25pt {\hfill (\Alph{section}\arabic{equation})%
\break}\label{A9}

\subsection{Short-distance expansions}

\[
\displaystyle\frac \partial {{\partial }p_\mu }  [\displaystyle\frac{%
\sin(p\epsilon )}p] = p_\mu \left[ -\displaystyle\frac{{\epsilon }^3}3 + %
\displaystyle\frac{p^2{\epsilon }^5}{30}  -\displaystyle\frac{p^4{\epsilon }%
^7}{30}+{\cal O}({\epsilon })\right]. 
\]
\stepcounter{equation}\vskip -25pt {\hfill (\Alph{section}\arabic{equation})%
\break}\label{A10}

\begin{eqnarray*}
\displaystyle\frac{\partial ^2}{{\partial }p_\mu {\partial }p_\nu }\left[ 
\frac{sin(p\epsilon )}p\right] &=& {\delta }_{\mu \nu }\left[ -\displaystyle%
\frac{{\epsilon }^3}3 +\displaystyle\frac{p^2{\epsilon }^5}{30} -%
\displaystyle\frac{p^4{\epsilon }^7}{30} +{\cal O}({\epsilon })\right] 
\nonumber \\
&& -p_\mu p_\nu \left[ -\displaystyle\frac{{\epsilon }^5}{15} -\displaystyle%
\frac{p^2{\epsilon }^7}{210} +\displaystyle\frac{p^4{\epsilon }^9}{7560}+%
{\cal O}({\epsilon })\right].
\end{eqnarray*}
\stepcounter{equation}\vskip -25pt {\hfill (\Alph{section}\arabic{equation})%
\break}\label{A11}

\eject


\begin{figure}
\centering
\input FEYNMAN
\begin{picture}(45000,500)
\drawline\gluon[\E\REG](0,0)[6]
\put(\gluonfrontx,\gluonfronty){$A$}
\put(\gluonbackx,\gluonbacky){$A$}
\drawline\scalar[\E\REG](8000,0)[4]
\put(\scalarfrontx,\scalarfronty){$c$}
\put(\scalarbackx,\scalarbacky){$\bar{c}$}
\drawline\gluon[\E\REG](20000,0)[3]
\put(\gluonfrontx,\gluonfronty){$A$}
\drawline\photon[\E\REG](\gluonbackx,0)[3]
\put(\photonbackx,\photonbacky){$B$}
\drawline\photon[\E\REG](33000,0)[6]
\put(\photonfrontx,\photonfronty){$B$}
\put(\photonbackx,\photonbacky){$B$}
\end{picture}

\begin{picture}(45000,20000)
\drawvertex\gluon[\S 3](200,10000)[4]
\drawvertex\gluon[\SE 4](15000,10000)[3]
\drawline\gluon[\S\REG](35000,10000)[4]
\drawline\scalar[\SW\REG](\gluonbackx,\gluonbacky)[3]
\drawline\scalar[\SE\REG](\gluonbackx,\gluonbacky)[3]
\end{picture}

\caption{\protect\small  Feynman Rules}

\vspace{5mm}
\begin{picture}(10000,5000)
\drawline\scalar[\E\REG](0,0)[2]
\drawloop\gluon[\NE 3](\pbackx,\pbacky)
\drawline\scalar[\E\REG](\pbackx,\pbacky)[2]
\drawline\scalar[\W\REG](\pbackx,\pbacky)[4]
\end{picture}

\caption{\protect\small  Ghost self-energy}

\vspace{5mm}



\begin{picture}(40000,10000)

\drawloop\gluon[\NE 0](10000,5000)
\drawline\gluon[\E\CENTRAL](\loopbackx,\loopbacky)[5]
\drawline\gluon[\W\FLIPPEDCENTRAL](\loopfrontx,\loopfronty)[5]
\put(10000,200){$(a)$}

\drawline\gluon[\E\REG](25000,5000)[2]
\drawline\scalar[\NE\REG](\gluonbackx,\gluonbacky)[2]
\drawline\scalar[\SE\REG](\scalarbackx,\scalarbacky)[2]
\drawline\gluon[\E\REG](\scalarbackx,\scalarbacky)[2]
\drawline\scalar[\SW\REG](\gluonfrontx,\gluonfronty)[2]
\drawline\scalar[\NW\REG](\scalarbackx,\scalarbacky)[2]
\put(25000,200){$(b)$}
\end{picture}

\caption{\protect\small Vacuum polarization tensor}


\vspace{5mm}
\begin{picture}(40000,20000)
\drawline\gluon[\S\REG](3000,10000)[2]
\drawline\scalar[\SW\REG](\gluonbackx,\gluonbacky)[3]
\drawline\scalar[\W\REG](\scalarbackx,\scalarbacky)[2]
\drawline\scalar[\SE\REG](\gluonbackx,\gluonbacky)[3]
\drawline\scalar[\E\REG](\scalarbackx,\scalarbacky)[2]
\drawline\gluon[\W\REG](\scalarfrontx,\scalarfronty)[8]
\put(3000,200){$(a)$}

\drawvertex\gluon[\S 3](25000,11000)[3]
\drawline\scalar[\W\REG](\vertexthreex,\vertexthreey)[2]
\drawline\scalar[\E\REG](\vertexthreex,\vertexthreey)[3]
\drawline\scalar[\E\REG](\scalarbackx,\scalarbacky)[2]
\put(25000,200){$(b)$}
\end{picture}
\caption{\protect\small  One-loop ghost-gluon vertex}
\end{figure}

\vspace{1cm}

\begin{thebibliography}{99}

\bibitem{fjl}  D.Z. Freedman, K. Johnson and J.I. Latorre, Nucl. Phys. {\bf %
B371} (1992) 353.\\ D. Z. Freedman, ``{\it Differential Regularization and
Renormalization: Recent Progress}'' in {\it Proceedings of the Stone Brook
Conference on String and Symmetries}.

\bibitem{hala}  P.E. Haagensen and J.I. Latorre, Phys. Lett. {\bf B283}
(1992) 293.

\bibitem{haa}  P.E. Haagensen, Mod. Phys. Lett. {\bf A7} (1992) 893.

\bibitem{hl}  P.E. Haagensen and J.I. Latorre, Ann. Phys. {\bf 221} (1993)
77.

\bibitem{ramon}  R. Munoz-Tapia, Phys. Lett. {\bf B295} (1992) 95.

\bibitem{dn}  G. Dunne and N. Rius, Phys. Lett. {\bf B293} (1992) 367.

\bibitem{fjmv}  D.Z. Freedman, K. Johnson, R. Munoz-Tapia and X.
Vilasis-Cardona, Nucl. Phys. {\bf B395} (1993) 454.

\bibitem{djt}  S. Deser, R. Jackiw and S. Templeton, Phys. Rev. Lett. {\bf 48%
} (1982) 975; Phys. Rev. {\bf D23} (1981) 2291; Ann. Phys. {\bf 140} (1982)
372; Ann. Phys. {\bf 185} (1988) 406.

\bibitem{prao}  R.D. Pisarski and S. Rao, Phys. Rev. {\bf D32} (1985) 2081;
E.R. Bezerra De Mello, Ann. Phys. {\bf 185} (1988) 401.

\bibitem{alr}  L. Alvarez-Gaume, J.M.F. Labastida and A.V. Ramallo,
Nucl.Phys. {\bf B334} (1990) 103.

\bibitem{gmr1}  G. Giavarini, C.P. Martin and F. Ruiz. Ruiz. Nucl. Phys. 
{\bf B381} (1992) 222; \\ C.P. Martin, Phys. Lett. {\bf B241} (1990) 513.

\bibitem{csw}  W. Chen, G.W. Semenoff and Y.S. Wu. Mod. Phy. Lett. {\bf A5}
(1990) 1833; Phys. Rev. {\bf D46} (1992) 5521.

\bibitem{cz}  W.F. Chen and Z.Y. Zhu, J. Phys. A: Math. Gen. {\bf 27} (1994)
1781.

\bibitem{brt1}  D. Birmingham, M. Rakowsky and G. Thompson, Phys. Lett. {\bf %
B251} (1990) 121; E. Witten, Comm. Math. Phys. {\bf 121} (1989) 351.

\bibitem{asfa}  M. Asorey and F. Falceto, Phys. Lett. {\bf B241} (1990) 31.

\bibitem{gmm}  E. Guadagnini, M. Martellini and M. Mintchev, Phys. Lett. 
{\bf B227} (1989) 111.

\bibitem{kll}  H.C. Kao, K. Lee and T. Lee, Phys. Lett. {\bf B373} (1996) 94.

\bibitem{singer} S. Axelrod and I.M. Singer, {\it Chern-Simons Perturbation
Theory}, In $*$ New York 1991, Proceedings of Differential Geormetric Method
in Theoretical Physics, {\bf 1} 3; {\bf hep-th/9110056}.\\
S. Axelrod and I.M. Singer, {\it Chern-Simons Perturbation Theory} {\bf II}, 
{\bf hep-th/9304987}.

\bibitem{asfa1} M. Asorey, I. Falceto, J.L. L\'opez and G. Luz\'on, 
Phys. Rev. {\bf D49} (1993) 5377; Nucl. Phys. {\bf B429} (1994), 344.

\bibitem{gmr2}  G. Giavarini, C.P. Martin and F. Ruiz Ruiz, NIKHEF-H 93-05, 
{\it Physical meaning and not so meaningful symmetries in Chern-Simons field
theory} (To appear in Phys. Rev. D).\\
G. Giavarini, LPTHE 92-42, ``{\it Low
dimensional topology and quantum field theory}'', (talk given at the NATO
AWR, 6-13 September, 1992, Cambridge (UK)); NIKHEF-H 93-05, ``{\it The
universality of the shift of the Chern-Simons parameter for a General Class
of BRST invariant regularization}''.

\bibitem{bs}  N.N. Bogoliubov and D.V. Shirkov, ``{\it Quantum Fields}''
(Benjamin, Reading, 1983).

\bibitem{sm}  V.A. Smirnov, Nucl. Phys. {\bf B427} (1994) 325; Z. Phys.{\bf C67},
 (1995) 531.

\bibitem{brt2}  D. Birmingham, M. Rakowsky and G. Thompson, Nucl. Phys. {\bf %
B234} (1990) 367; F. Delduc, F. Gieres and S.P. Sorella, Phys. Lett. {\bf %
B225} (1989) 367; P.H. Damgarrd and V.O. Rivelles, Phys. Lett. {\bf B245}
(1989) 48.

\bibitem{dor}  D. Daniel and N. Dorey, Phys. Lett. {\bf B246} (1990) 82; N.
Dorey, Phys. Lett. {\bf B 246} (1990) 87; F. Delduc, C. Lucchesi, O. Piguet
and S.P. Sorella, Nucl. Phys. {\bf B346} (1990) 313.


\bibitem{lee}  B.W. Lee and J. Zinn-Justin, Phys. Rev. {\bf D5} (1972) 3121.\\
J.C. Collins, ``{\it Renormalization}'' (Cambridge University
Press, Cambridge, 1987). 
\end{thebibliography}
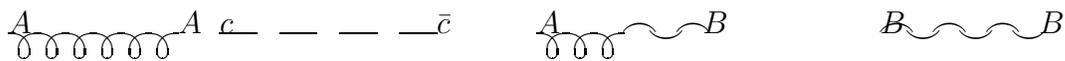
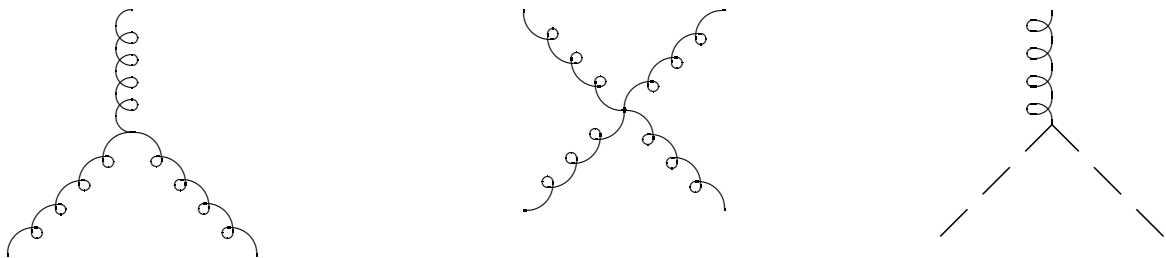
\end{document}